\documentclass[a4paper,10pt]{article}
\usepackage{graphicx}
\usepackage{latexsym}

\addtocounter{page}{-1}
\begin{document}
\pagestyle{empty}
\parindent 8mm
\pagestyle{empty}

\vspace*{2cm}
\title{Simulation of the weighted spatio-structured three cornered game and its satatistical properties}
\author{Shota Hayakawa and Norihito Toyota}
\date{Hokkaido Information University,
59-2 Nishinopporo Ebetsu City \\
E-mail:toyota@do-johodai.ac.jp
}

\maketitle
\baselineskip 5mm
\pagestyle{empty}
\begin{abstract}
In this article, we study the evolution of the strategies in a sort of weighted three cornered game, Glico Game. 
This game has three strategies in each agent and three parameters in the payoff matrix.  
 We divide the parameter region into 8 areas and explore its qualitative and quantitative properties of the evolution of strategies which include Haming distance, correlation functions and the population sequence of each strategy.  
When the three payoff parameters take nearly values each other, the complex patterns mainly appear and as the difference of parameter values are enlarged, some peculiar strategy trends to be dominant gradually.  An interesting fact is that the strongest strategy in the payoff do not necessarily dominate the latice world. There are even the cases where the weakest strategy in the payof dominate. Generally many cases have some period much shoter than the dimension of the configuration space. 
\end{abstract}
\bf{key words:} 
\it{weighted three-cornerd game, spatio structured, itareted game}

\rm
\normalsize
\baselineskip 6mm
\pagestyle{empty}
\thispagestyle{empty}

\renewcommand{\thefootnote}{\fnsymbol{footnote}} 
\rm
\normalsize
\baselineskip 6mm

\section{Introduction}

The descrete dynamics is useful to understand various phenomena in the natural and artificial world which includes biological, social, ecological phenomena and so on.
 The game theory among many approaches give one of interesting view of them. 
Most of one shot games  can be analyzed analytically \cite{Okad}.   
Many researchs have been made to study the Dilemma, pareto optimal, Nash equilibrium, and how can we obtain cooperation in the Dilemma games rationally.  
Based on Prisoner's Dilemma (PD), iterated games have been intensively studied and Axelrod \cite{Axel} has shown that cooperation can emerge as a norm in a society comprised of individuals with selfish motives. 
Moreover spatial version of PG is discussed so that a defection is to be only evolutionary stable strategy (ESS) \cite{Smit} if each agent interacts with any other agents. 
This aspect drastically changes if a spatial structure of the population is considered. If the interaction between agents is locally restricted to their neighbors\cite{Nowa}, a stable coexistence between cooperators and defectors become possible under certain conditions\cite{Nowa}\cite{Nowak}. 
In the case, one agent plays PD with their neighbors and in next step the agent take the same strategy of the agent that acquired highest payoff among the neighbors,  which reflects to Dawin's theory. 
It is assumed that all agents play at same time and follow the same way. 
Recently they are systematically have been disucussed \cite{Schw}.  

In this current of research, we discuss a spatio evolutionary game with three strategies, which has not been explored so heavily \cite{Szab}.
Now assume there are two persons (countries and so  on). 
One person A is postulated to be stronger than another person B in a game. In the game A necessarily wins over B. 
That is trivial.  If an outsiders participates in the game, the situation, however, would become dificult. 
Especially when the outsider C is stronger in the game than A but weeker than B, the situation becomes a three cornered game.  
When a super country bullys a puny country, some outsider with which the super country does not like to fight and which want to obtain some kind of interests from the puny  country, intrade on them. 
Such a situation often may arise. 
The situation which we consider is that there are many countries in such relation and they often repeat the interactions. This primafacie symmetric setup brings some drastic results. 
The aim of this article is to uncover such aspects by excuting a computer simulation.   

The plan of this article is as follows; The next section 2, we describe qualitative aspect of simulation results for eight cases. The quantitative analyses are obtained. The last section is devoted to summary. 

\section{Weighted Three Cornerd Game}

We consider the paper-rock-scissors as a metaphor of the three cornered game. 
Since the game is completely symmetric, any outcome will express symmetric ones. 
So we introduce asymmetric payoff for each strategy which is called "Glico Game" (G-game) in Japan. 
Let prepare three strategies, R, G and B. 
We assume that
\begin{center}
R wins G and get the payoff $P_R$,\\ 
G wins B and get the payoff $P_G$, \\
B wins R and get the payoff $P_B$. 
\end{center}
We explore the evolution of the population of the strategirs, changing the parameters, $P_R, \; P_G,\; P_B$. 
Essentially there are following eight patterns in the parameter space, considering symmetry in the strategies;\\
(1)$P_R > P_G > P_B$ and the differences of the payoffs are all  small.\\
 (2)$P_R > P_B > P_G$ and the differences of the payoffs are all  small.\\
(3)$P_R >> P_G > P_B$ and the differences of $P_R - P_B$ only is  small.\\
(4)$P_R >> P_B > P_G$ and the differences of $P_R - P_G$ only is  small.\\
(5)$P_R > P_G >> P_B$ and the differences of $P_G - P_R$ only is  small.\\
(6)$P_R > P_B >> P_G$ and the differences of $P_B - P_G$ only is  small.\\
(7)$P_R >> P_G >> P_B$ and the differences of the payoff are all large.\\
(8)$P_R >> P_B >> P_G$ and the differences of the payoff are all large.\\
Without losing generality we can choose $P_R$ as the largest payoff.
In the cases with odd number in the above the order of strength of the strategies is consistent with the order of the payoff and the cases with even nunber are not so. 
We call this weighted three cornered game with general payoffs G-game.  
Then the payoffs are obtained in the table 1.

\begin{table}[h]\centering
\caption{Payoff table for G-game. The left and right variables in the parenthesis show the payoffs acquired by the agent $i$ and $j$, respectively. }
\begin{tabular}{|c|c|c|c|} \hline
  &\makebox[15mm]{R$_j$} & \makebox[15mm]{G$_{j}$} & \makebox[15mm]{B$_{j}$} \\ \hline
R$_{i}$ & ($0,0$) & ($P_R,0$) & ($0,P_B$)\\ \hline
G$_{i}$ & ($0,P_R$) & ($0,0$) & ($P_G,0$)\\ \hline
B$_{i}$ & ($P_B,0$) & ($0,P_G$) & ($0,0$)\\ \hline
\end{tabular}
\end{table}

We make this setting a spatio-evolutionary game. 
We consider $n \times n=N$ lattice-like cells on which agents are lying with the strategies. 
The agents play the G-game with the Moor neightborhood and theirself with a fixed strategy.   
In next round the agents take the same strategy as that of the player who has gotten highest payoff among the neighborhood and theirself in the present round. 
 We simulate it by $t$ rounds to investigate the evolution patterns of this game. 

\section{Clussification and Simulation Results}
\subsection{Qualitative characters}
  In the simulation we take $n=51$ and the iteration number $t$ up to convergence (maximal 1000 rounds). 
Black cells show the Red strategy, the gray cells show the Green strategy and the white cells show the Blue strategy  in figures. 
The periodic boundary condation is taken on the latice and an random initial configuration with equi-probabilities for the strategies R, G and B are taken in the all simulations.  \\

(1)Case with $P_R =1.03 > P_G =1.02 > P_B =1.01$\\
First R is dominant and then temporally B is dominant,  because B can exploit  R that is strongest strategy in the payoff. 
After that, the configuration traces some complicate histries to mainly reach three patterns.  
They are  square spiral patterns (Fig. 4), the patterns with multiple square and intricate patterns (Fig.1) and we call them complex patterns, generically. 
Up to 400 rounds, they show periodic behabiors (Fig.1 and Fig.2) or irregular pattern with equi-probability. 
However, the formers appears twice as often as the latters up to 1000 rounds. 
This means that though many cases do not converge at 400 rounds, they would finally converge to some patterns. 
Globally  G has the largest population and occupies nearly 40 percent of the whole, and the remainders are occupied by R and B with nearly equal ratio (Fig.1).  

 R (or B or G) dominant configurations with some B-G poles (or R-G  or R-B poles)  sometimes appear. Their total frequency is about a quater of all simulations. While R dominant configuration appears after a few dozen, B and G dominant confiruration appear through long complex patterns, eventually. 
The poles circulates through whole lattice world to return again (refer to Fig.5).  \\ 

%

\begin{figure}[htbp]
\begin{center}

\includegraphics[width=\linewidth]{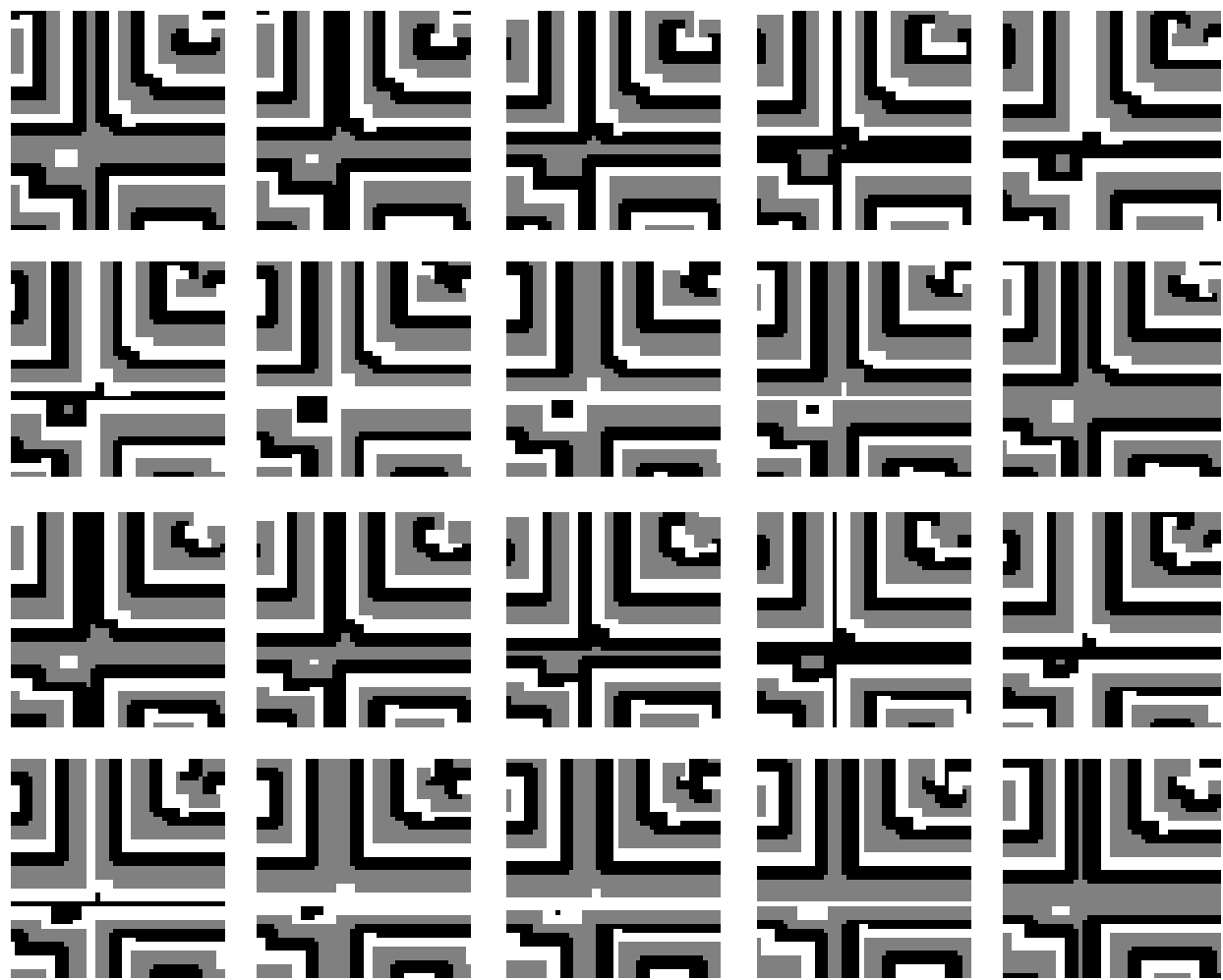}

\caption{Typical multiple square pattern at $ 71 \sim 90 $ steps.}
\end{center}
\end{figure}

(2)$P_R= 1.03 > P_B=1.02 > P_G=1.01$ \\
First B is dominant and then temporally G is dominant,  because G can exploit B. In these cases the result is similar to (1) except that there are no R (or B or G) dominant configurations with some pole like objects. 
While the periodic complex patterns appears with a frequency quadruple to that of the irregularly intricate pattern (Fig.3), the 1000 round simulations reverse the frequency rate between both patterns. In general the cases in (2) converge more slowly than the cases (1). 
 R, G and B occur with nearly equal possibility.   \\    


\begin{figure}[tbp]
\begin{center}

\includegraphics[width=\linewidth]{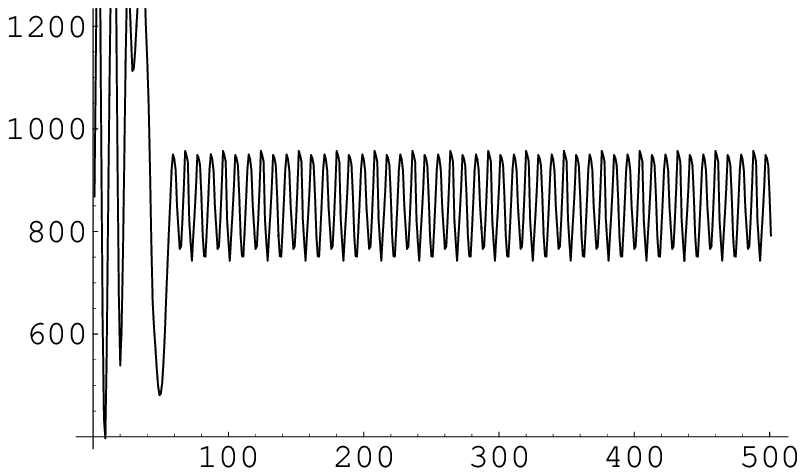}
\includegraphics[width=\linewidth]{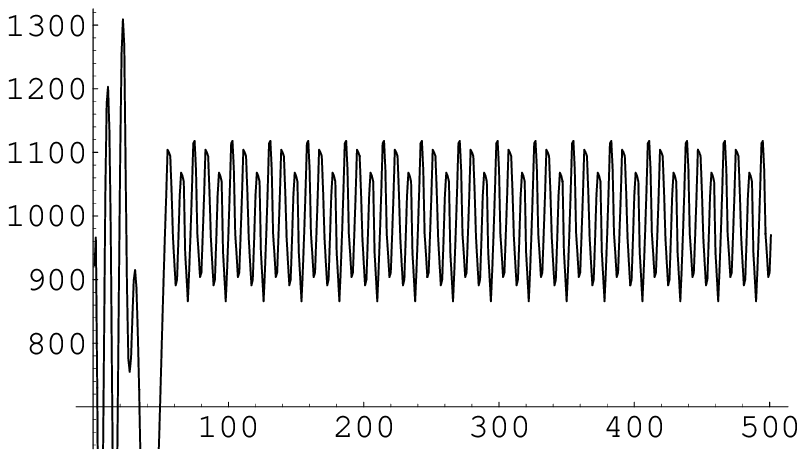}
\includegraphics[width=\linewidth]{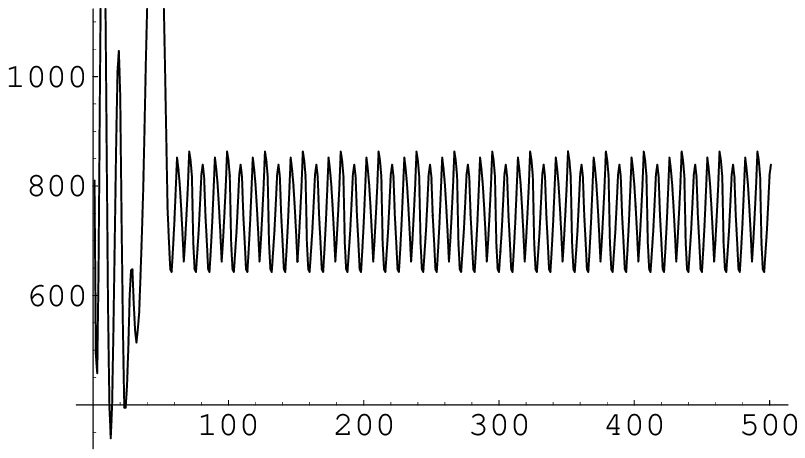}

\caption{The population sequence of R, G and B in fig.1.}


\includegraphics[width=\linewidth]{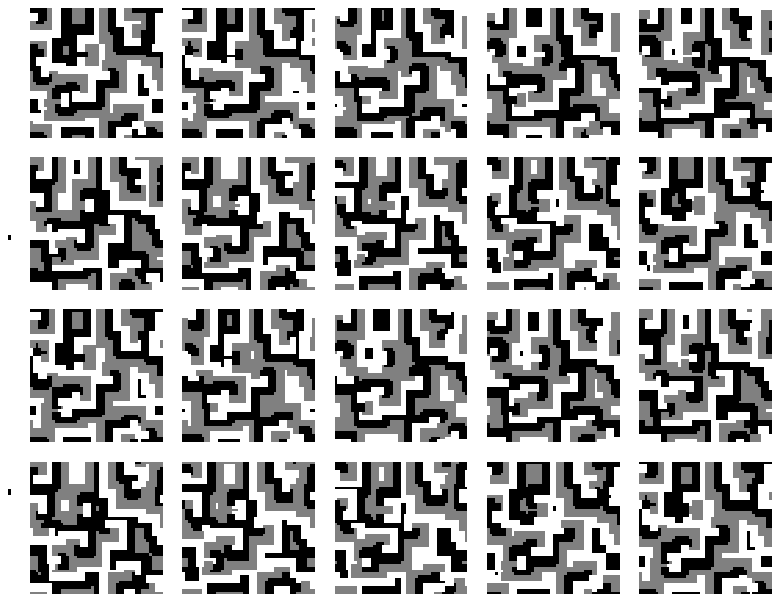}

\caption{Typical irregularly intricate pattern at $ 131 \sim 150 $ steps.}
\end{center}
\end{figure}

(3)$P_R =2>> P_G =1.02 > P_B =1.01$\\
The initial histories of configurations are mainly R$\rightarrow$B, and sometimes  R$\rightarrow$B$\rightarrow$R($\rightarrow$G). 
The periodic complex patterns appear with about 2/3 of the possibility (Fig.4).

  The remnants of them are the complete R dominant configuration and the complete B one with about 1/6 of the probability, respectively. 
In this case the second strong strategy G in the payoff never becomes a dominant strategy.  
 The distribution ratios of R, G and B are approximately 4:3:1. 
\\

(4)$P_R =2>> P_B=1.02 > P_G=1.01$ \\
These cases converge with a short time,  comparatively. 
There are no cases with irregularly intricate patterns. 
They almost converge to the complex patterns with 80 percent of the probability  through R$\rightarrow$B (sometimes $\rightarrow$G), and most of others finish at the complete B configuration, which is the second strong strategy in the payoff,  through R$\rightarrow$B. 
For the case corresponding to that through R$\rightarrow$B $\rightarrow$G, 
a complete G configulation appears, ralely. 
The strongest strategy R in the payoff never appears. 
The distribution ratios of R, G and B are approximately same as (3), 4:3:1.\\


(5)$P_R =2.01> P_G =2 >> P_B =1.0 $ \\
Depending on initial configurations, the various patterns (almost all patterns   in the present simulations) appear in these cases. 
The complete B configuration (sometimes with R-G fractions or poles) and 
the complete R configuration (sometimes with G-B fractions or poles) appear with about 25 percent and about 70 percent, respectively (see Fig. 5 and 6). 
In the cases with  the thick (G-B) poles, the patterns look like striped patterns, which occupy a few percent of the whole.    
The second strong strategy G in the payoff never becomes dominant like (3). 
The complex patterns sometimes appears, too.\\


\begin{figure}[htbp]
\begin{center}

\includegraphics[width=\linewidth]{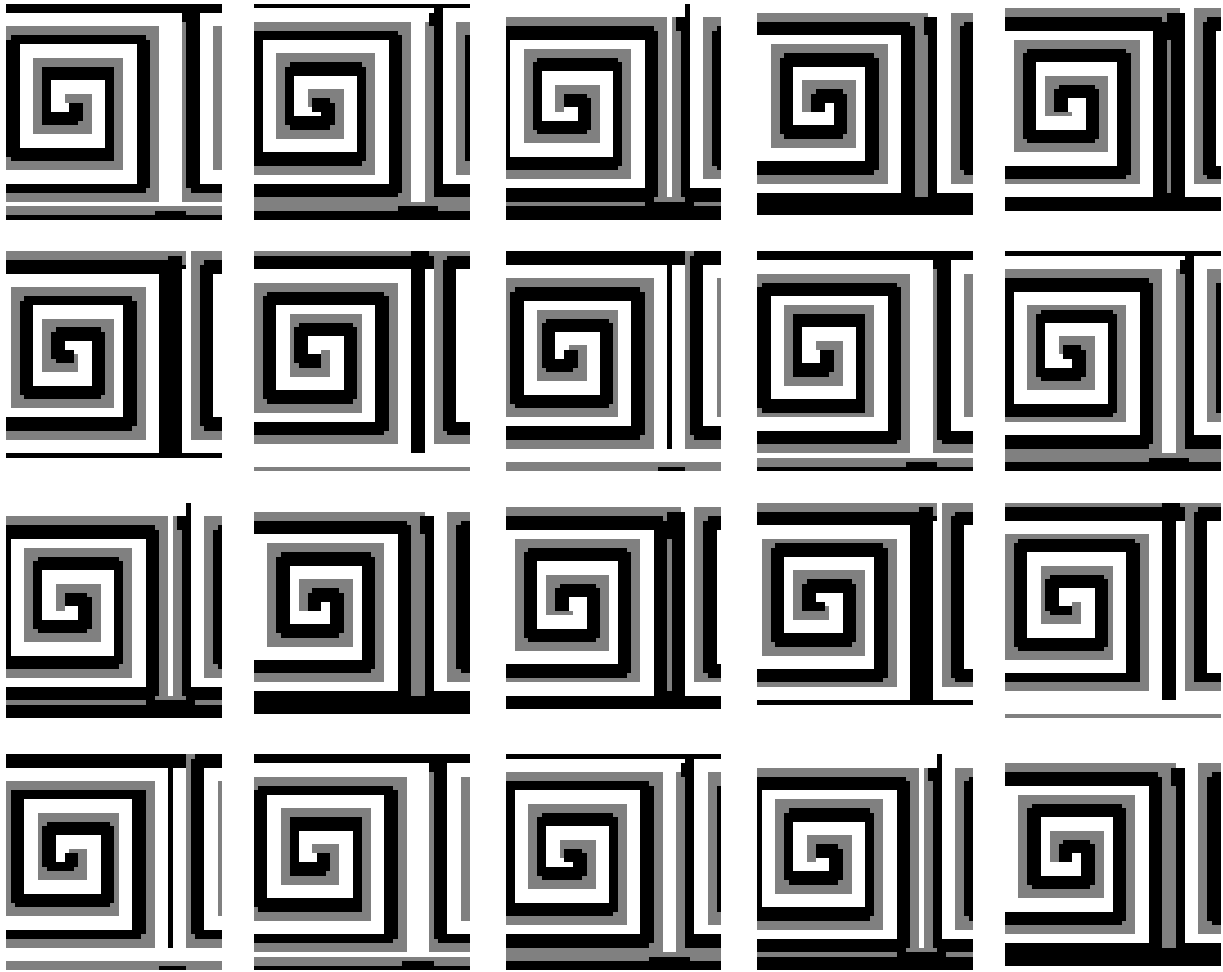}

\caption{Typical irregularly intricate pattern at $ 41 \sim 60 $ steps.}
\end{center}
\end{figure}

\begin{figure}[htbp]
\begin{center}

\includegraphics[width=\linewidth]{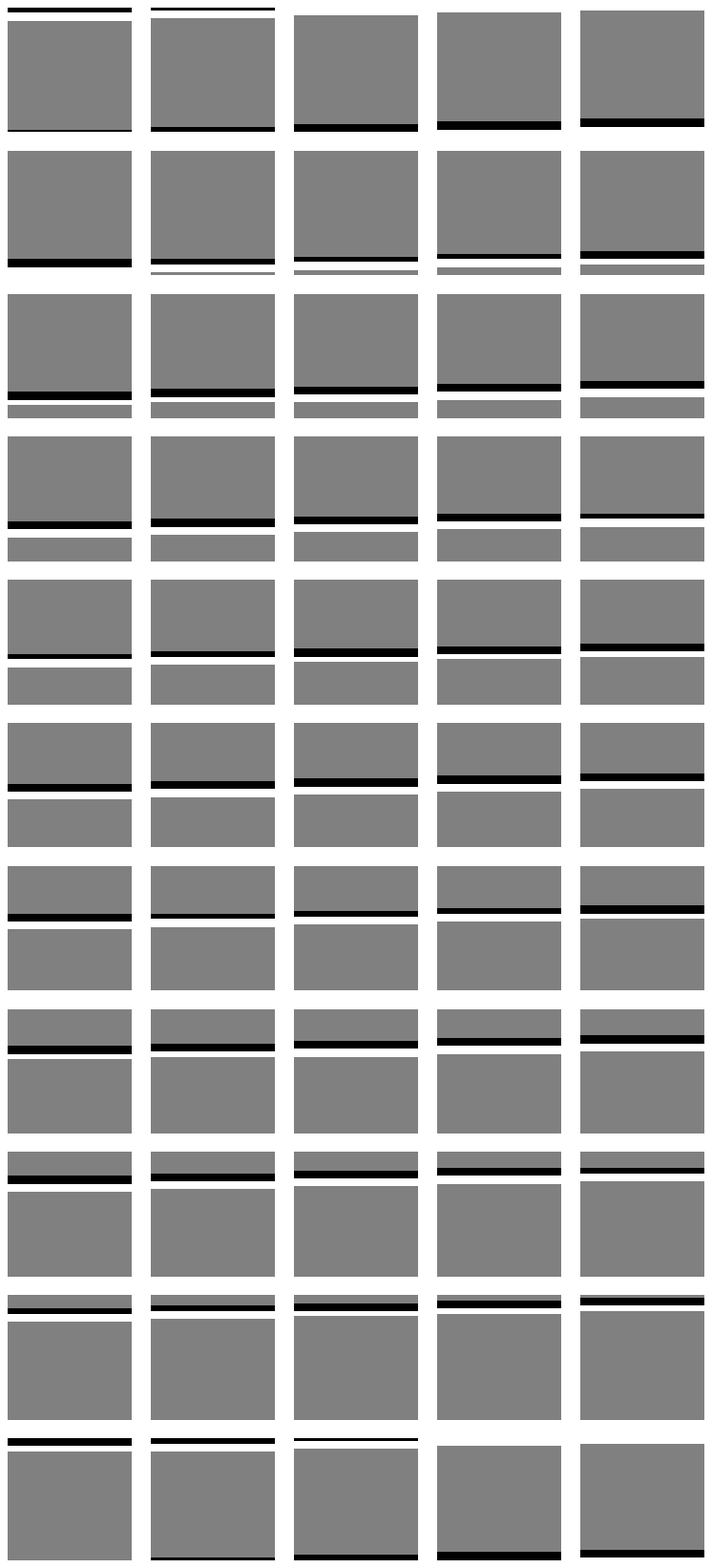}

\caption{Typical pattern with a pole at $ 51 \sim 100 $ steps.}
\end{center}
\end{figure}

\begin{figure}[htbp]
\begin{center}

\includegraphics[width=\linewidth]{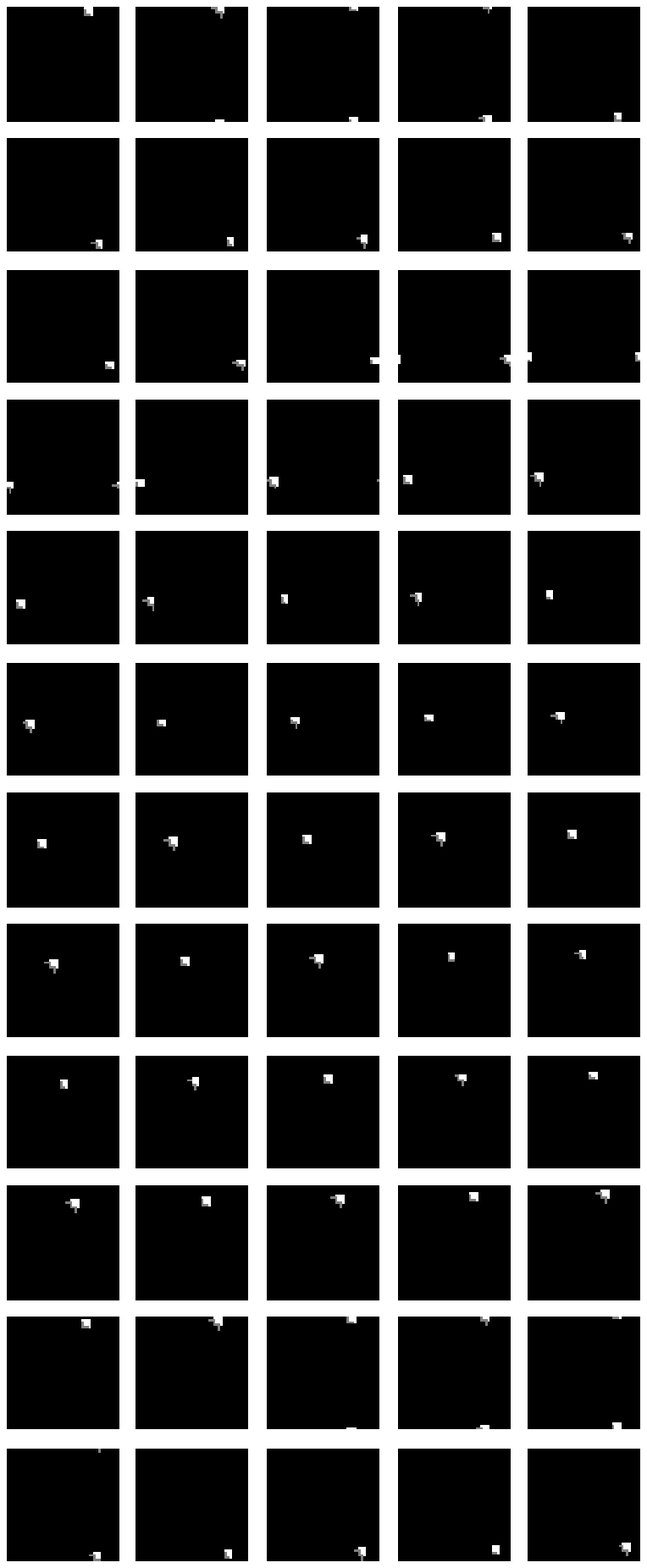}

\caption{Typical pattern with a fraction at $ 41 \sim 100 $ steps.}
\end{center}
\end{figure}

(6)$P_R =2.01> P_B=2 >> P_G=1$ \\
In these cases the weakest strategy G in the payoff almost comes to be dominant. The complete G appears with about  25 percent of the probability and G dominant with some R-B poles or fractions configuration appear with about 60 percent, and their total is about 85 percent. 
They go through B $\rightarrow $G at early period.  
The patterns with thick poles aresed with a few percent look like striped object.
 Ralely hightly irregular patterns appear.  \\
 
(7)$P_R =3 >> P_G =2>> P_B=1$ \\
The strongest strategy R in the payoff completely cover the world. This aspect is stable against the variance of initial states. As an extreme rare case, a complete B configuration appears.  \\

(8)$P_R =3 >> P_B=2 >> P_G=1$ \\
The history of dominant strategy is such as $R \rightarrow B \rightarrow G$. 
 In maked contrast with the case (7) the weakest strategy G in the payoff completely cover the world with about 60 percent or becomes dominant with a periodic pole with about 15 percent. 
The histories are sometimes cut off to be the complete  B or sometimes with R-G poles which lead to  B dominant configurations. 
These are summarized in the table 2.\\ 
\begin{table*}[t]
\caption{Quaritative summary in the eight kinds of parameter set. }
\begin{tabular}{|l|l|l|c|c|c|}\hline 
&\makebox[5mm]{$P_R$} &\makebox[5mm]{$P_G$}&\makebox[5mm]{$P_B$}
&\makebox[20mm]{early history of strategies} &\makebox[35mm]{mainly final states} 
\\ \hline \hline 
(1)& 1.03&1.02&1.01& R$\rightarrow$ B& periodic complex,(irregular ?), R-,G-,B-dominant
\\ \hline
(2)& 1.03&1.01&1.02&B $\rightarrow $ G& periodic complex (irregularly intricate ? )
\\ \hline
(3)& 2&1.02&1.01& R $\rightarrow$ B or R$\rightarrow$B $\rightarrow$R  ($\rightarrow$G) & periodic complex, sometimes all R or B 
\\ \hline
(4)& 2&1.01&1.02& R$\rightarrow$ B ($\rightarrow$ G)&  periodic square spiral, sometimes all B (ralely all G)
\\ \hline
(5)& 2.01&2&1& R (sometimes $\rightarrow$ B)& all R (+ poles), all B (+poles or fractions), complex, (ralely striped)
\\ \hline
(6)& 2.01&1&2& B$\rightarrow$ G& all G or almost G with fractions or poles, (ralely striped) 
\\ \hline
(7)& 3&2&1& R& all R
\\ \hline
(8)& 3&1&2& R$\rightarrow$ B ($\rightarrow$ G) & all G (+poles), sometimes all B (+ poles or fraction) 
\\ \hline
\end{tabular}
\end{table*}

First dominated strategy is almost the strongest one in the payoff. 
 (The exception happens in the cases (2) and (6) where the highest payoff $P_R$  nearly equal to the second high one, $P_B$.)  
This is a reasonable result but does not necessarily express the fact that it is the final dominant strategy.  
Usually the strategy is soon replaced by the predator strategy. 

The prey of the final dominant strategy does not carry the weakest strategy in the payoff but there are even the cases that  the weakest one or the strategy that has nearly equal payoff of the weakest strategy become the winner. 
This is an adventure and an interesting phenomenon. 

   Generally speaking, when three payoffs take nearly equal values such as in (1) and (2), the final population of each strategy takes nearly equal values. 
In fact the complex pattern mainly appear in only (1) $\sim$ (4).  
As the difference of them are enlarged, the difference of final populations 
increases so that the dominance of a specific strategy comes to be  peculiar.    


\subsection{Statistical Futures}
The complex patterns arised often in the cases (1) $\sim$ (5) have periodical structure. We observed it in the fig. 2.  
On the other hand the sequence of the populations often shows chaotic behabior in the irregularly  intricate patterns. 
Fig. 7 corresponds to the data of Fig. 3.
These non-periodic patterns arised in (1) and (2) may be presumed not to converge on some specific patterns.  
Though the periods of various cases are considerably long and about $500 \pm 400$ in the cases (1) and (2), the period runs from a few rounds to 200 rounds in (3) and (4).  

\begin{figure}[htbp]
\begin{center}

\includegraphics[width=\linewidth]{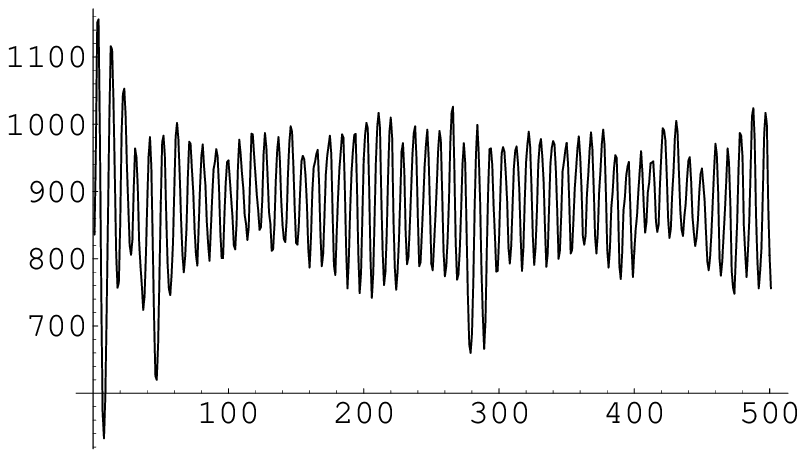}
\includegraphics[width=\linewidth]{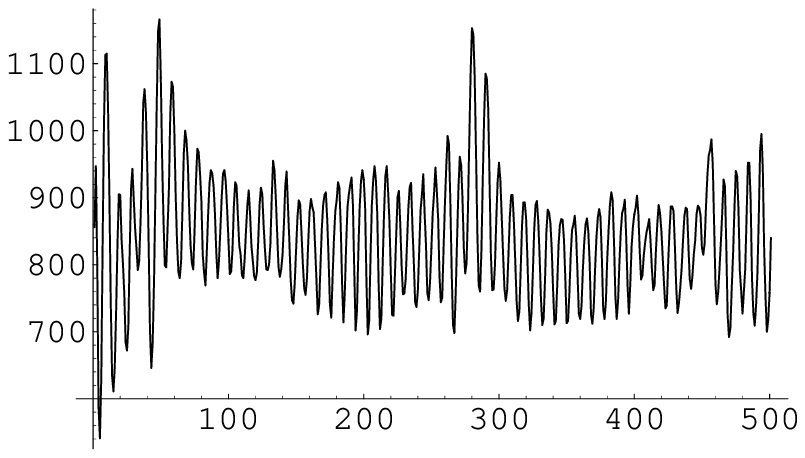}
\includegraphics[width=\linewidth]{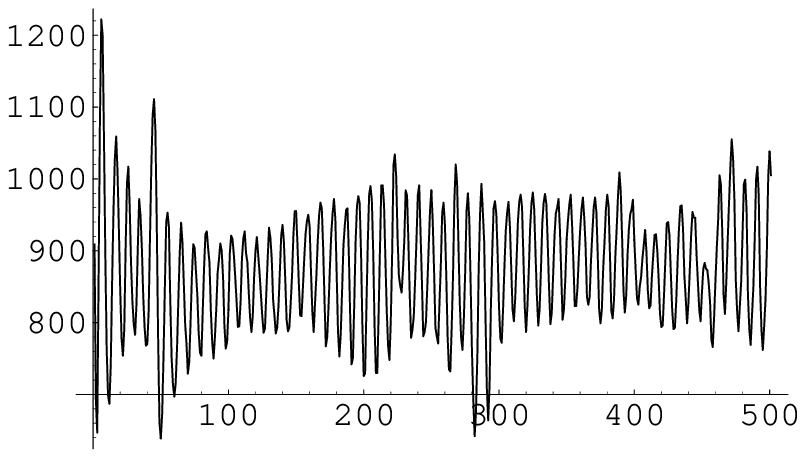}

\caption{The population change of R, G and B in the the irregularly  intricate pattern.}
\end{center}
\end{figure}

Almost cases  with some poles and/or fractions are also periodic  except the part of (6), where the sizes and the constituent ratios of poles or fractions  are invariant under the time transition but only the position of them moves 
on the latice (Fig.5, Fig.6, Fig. 8, Fig. 9 and Fig. 10). 
So we conjecture that except the cases with non-cyclic poles and fraction in (6) all cases are periodic or converge to one strategy. 
Of course all cases are in principle periodic on a finite latice. In the present cases the dimension of the configuration space is $3^{51\times 51}$ and so the fact that the period is $O(100)$ is highly nontrivial.

\begin{figure}[htbp]
\begin{center}

\includegraphics[width=\linewidth]{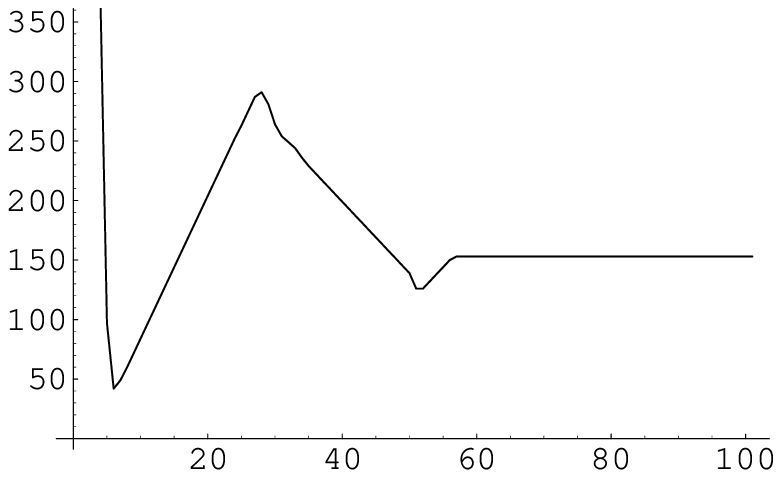}
\includegraphics[width=\linewidth]{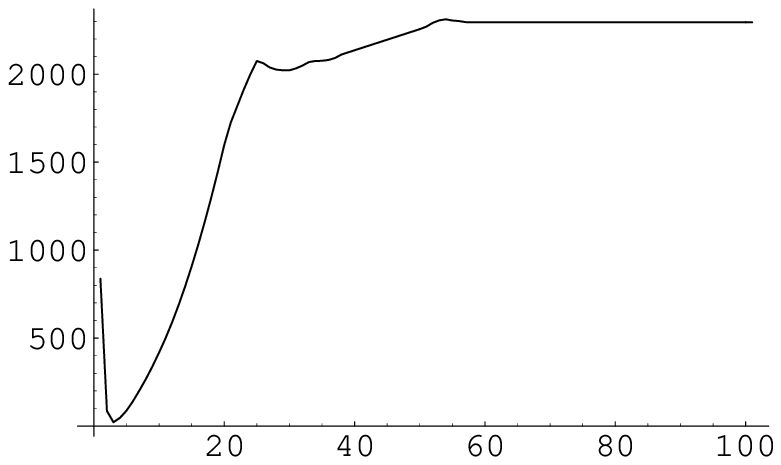}
\includegraphics[width=\linewidth]{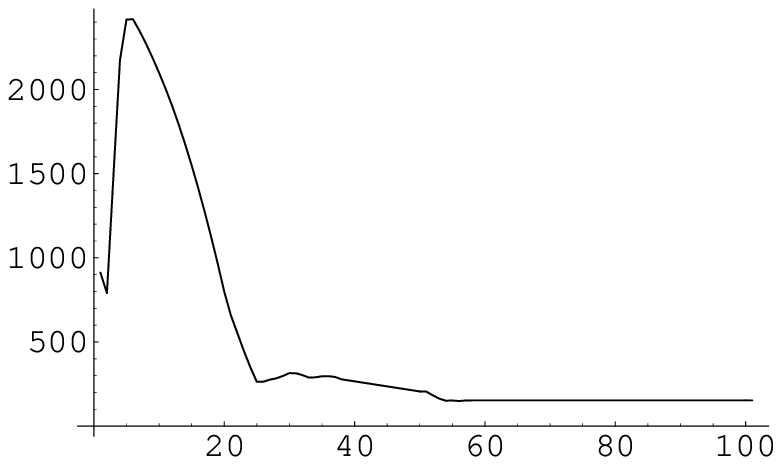}

\caption{The population change of R, G and B in a case with a pole, respectively. }
\end{center}
\end{figure}

\begin{figure}[htbp]
\begin{center}

\includegraphics[width=\linewidth]{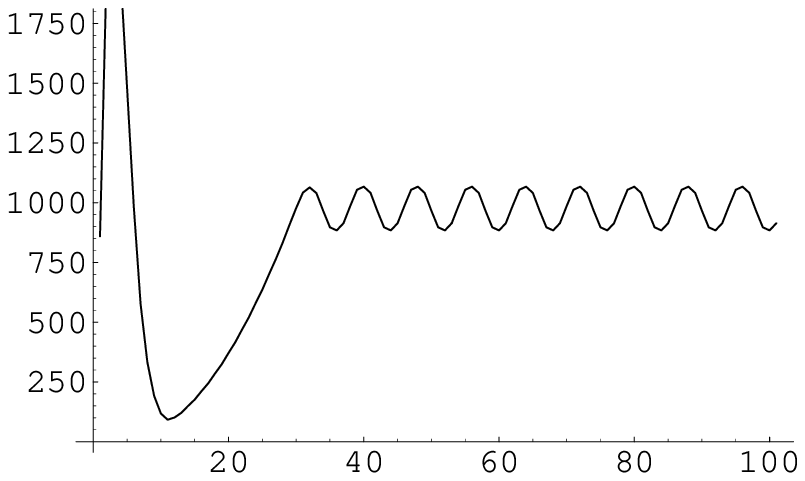}
\includegraphics[width=\linewidth]{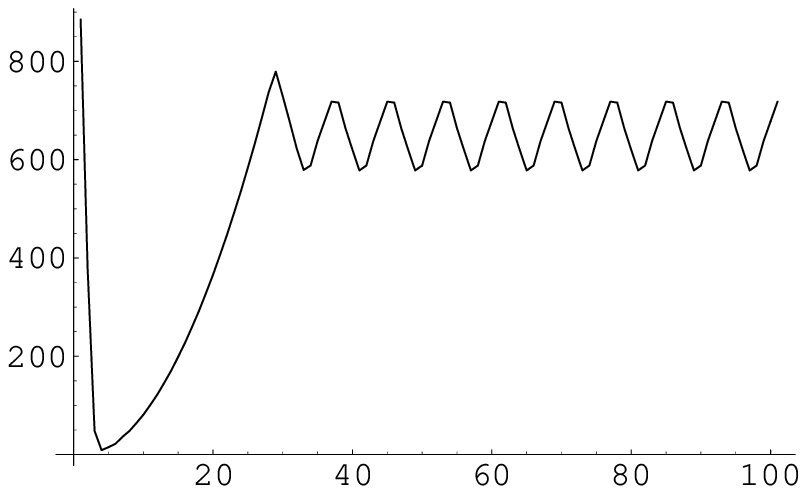}
\includegraphics[width=\linewidth]{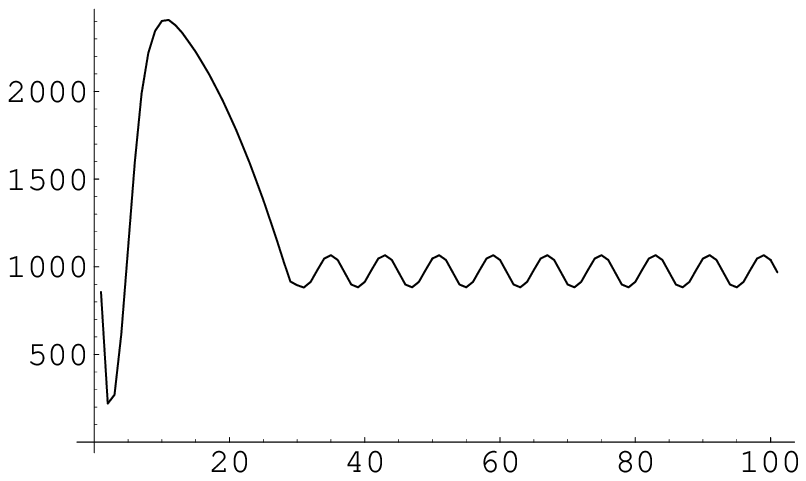}

\caption{The population change of R, G and B in a case with a fraction, respectively.}
\end{center}
\end{figure}

\begin{figure}[htbp]
\begin{center}

\includegraphics[width=\linewidth]{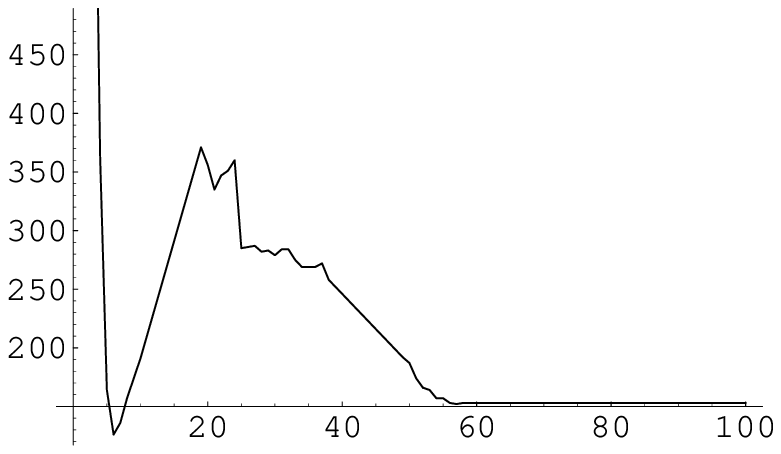}
\caption{The sequence of the Haming distance in the Fig.8. }



\includegraphics[width=\linewidth]{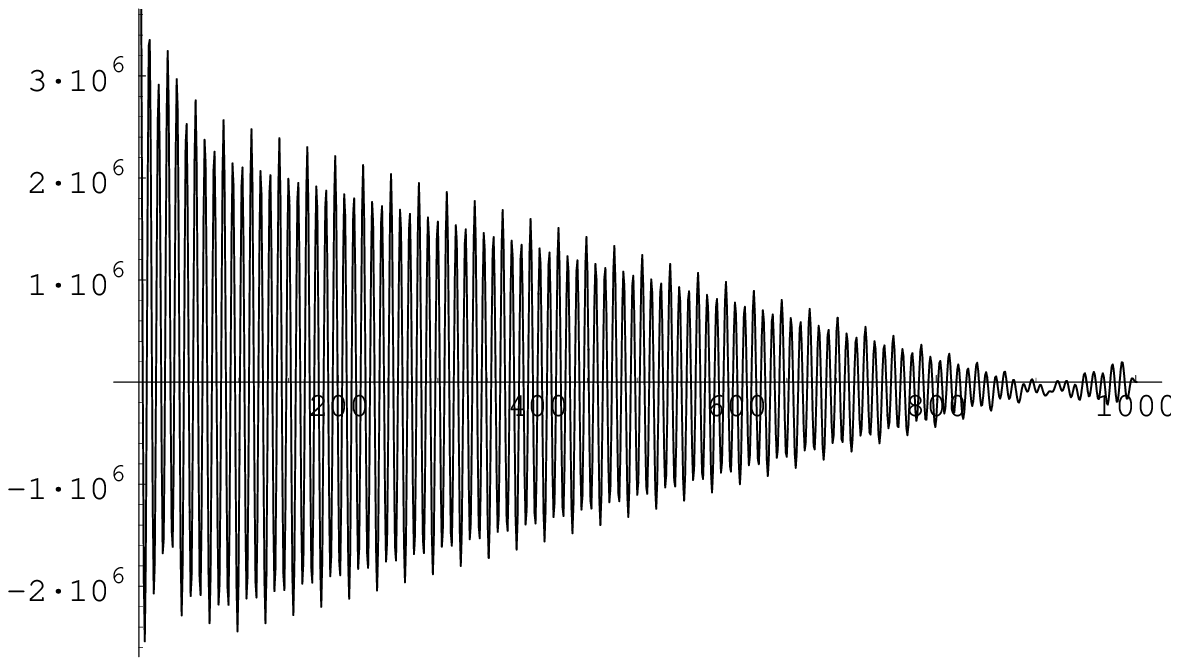}
\caption{The correration function of a periodic complex pattern. }



\includegraphics[width=\linewidth]{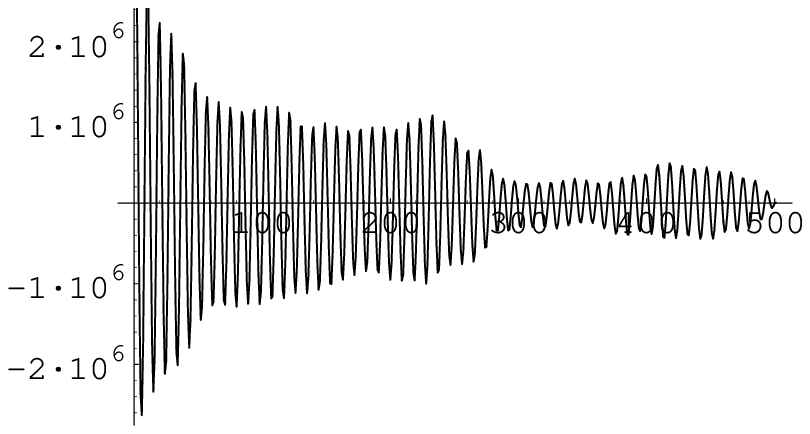}
\caption{The correration function of an irregularly intricate pattern. }



\includegraphics[width=\linewidth]{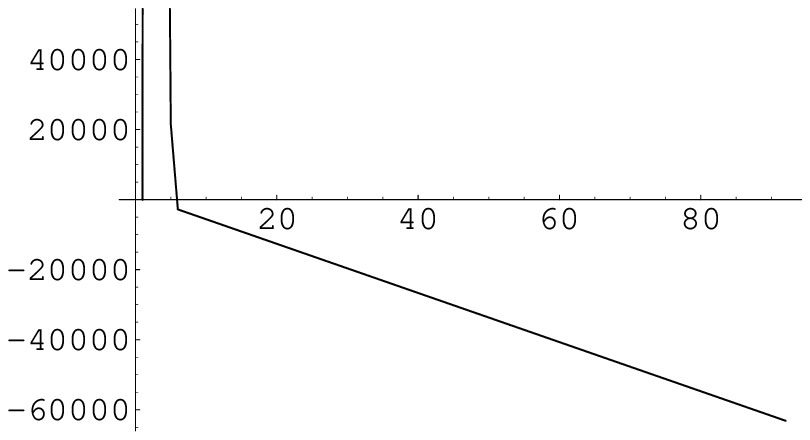}
\caption{The correration function of a R dominant pattern. }

\end{center}
\end{figure}

The Haming distance show the same behabior as the corresponding population 
sequence in essentials (Fig.10). 
The correlation functions monotonically decrease to zero with respect to 
the round number $t$ in the periodic complex patterns. 
 The function decrease, showing complicated behabior but to non zero values
 in the irregularly intricate cases, generally. 
When a specfic strategy is dominant,  the correlation function linearly decrease to large negative value. 
These are diagnostic features of correlation function (Fig.11, Fig. 12 and Fig 13).  They seem to be different from chaotic features.

\section{Concluding Remarks}
We have studied the evolution of the strategies in a sort of weighted three cornered game, Glico Game. 
Then we divide the parameter region with respect to  the payoff into 8 areas and explore its qualitative and quantitative properties.  
When the three payoff parameters take nearly values each other, the complex patterns mainly appear and as the difference of parameter values are enlarged, some peculiar strategy trends to be dominant gradually.  An interesting fact is that the strongest strategy in the payoff do not necessarily dominate the latice world. There are even the cases where the weakest strategy in the payof dominate. 
Generally many cases have some period much shoter than the dimension of the configuration space. 

Some irregular patterns maight converge to some specific patterns. 
One of our sequent studies is to uncover the character (chaotic, periodic, ...)  of the irregular patterns. 
It can be also expected that there are  some critical points and further detail investigations in each parameter region would are also needed.

\end{document}